\documentclass[amssymb,useAMS,prd,aps,amsmath,twocolumn,superscriptaddress,nofootinbib,preprintnumbers]{revtex4-1}
\pdfoutput=1

 \usepackage{hyperref}
 \hypersetup{
  colorlinks   = true, 
  urlcolor     = blue, 
  linkcolor    = blue, 
  citecolor   = red 
}
 \usepackage{slashed}

\usepackage[T1]{fontenc}
\usepackage{ae,aecompl}
\usepackage{soul}
\usepackage[usenames,dvipsnames]{color}

\usepackage{ulem}
\usepackage{graphicx}	
\usepackage{amsmath}	
\usepackage{amssymb}	
\usepackage{multirow}

\newcommand{\remove}[1]{}

\newcommand{\Eg}{E_\gamma}
\newcommand{\lel}{\lambda_e}
\newcommand{\lell}{\lambda_\ell}
\newcommand{\lga}{\lambda_\gamma}
\newcommand{\mmu}{m_\mu}
\newcommand{\mpi}{m_\pi}
\newcommand{\me}{m_e}
\newcommand{\Epsilon}{\mathcal{E}}

\DeclareUnicodeCharacter{2009}{\,} 

\begin{document}

\title{Using circular polarisation to test the composition and dynamics of \\ astrophysical particle accelerators}  

\author{C\'eline B\oe hm} 
\email{celine.boehm@sydney.edu.au}
\affiliation{School of Physics, Physics Road, The University of Sydney, NSW 2006 Camperdown, Sydney, Australia}
\affiliation{Institute for Particle Physics Phenomenology, Durham University, South Road, Durham, DH1 3LE, United Kingdom}
\affiliation{LAPTH, U. de Savoie, CNRS,  BP 110, 74941 Annecy-Le-Vieux, France}
\affiliation{Visiting Fellow, Perimeter Institute for Theoretical Physics, Waterloo ON N2L 2Y5, Canada}
\author{C\'eline Degrande}
\affiliation{Centre for Cosmology, Particle Physics and Phenomenology (CP3), Universit\'e catholique
de Louvain, B-1348 Louvain-la-Neuve, Belgium}

\author{Jakub Scholtz}
\affiliation{Institute for Particle Physics Phenomenology, Durham University, South Road, Durham, DH1 3LE, United Kingdom}

\author{Aaron C. Vincent}
\affiliation{Arthur B. McDonald Canadian Astroparticle Physics Research Institute, Department of Physics, Engineering Physics and Astronomy, Queen's University, Kingston ON K7L 3N6, Canada}
\affiliation{Visiting Fellow, Perimeter Institute for Theoretical Physics, Waterloo ON N2L 2Y5, Canada}

\begin{abstract}
We investigate the production of circularly polarised X and gamma-ray signals in  cosmic accelerators such as supernova remnants and AGN jets.  
Proton-proton and proton-photon collisions within these sites produce a charge asymmetry in the distribution of mesons and muons that eventually leads to a net circular polarisation signal as these particles decay radiatively. We find that  the fraction of circular polarisation thus produced is at the level of $ 5 \times 10^{-4}$, regardless of the exact beam spectrum, as long as it is made predominantly of protons. While this fraction is very small, the detection of circular polarisation signals  in conjunction with high-energy neutrinos would provide an unambiguous  signature of the presence of high-energy protons in cosmic accelerators. In Supernovae shocks in particular,  this would indicate the presence of relativistic protons  hitting stationary protons and/or low-energy photons in the intergalactic or interstellar medium. 
\end{abstract}
\maketitle

\section{introduction}
The observation of cosmic rays (CRs) over several orders of magnitude in energy indicate that particle acceleration occurs in many astrophysical sites and gives rise to high-energy particles which eventually escape and interact with the diffuse surrounding medium.  Supernova shocks, massive stars, pulsars, stellar OB associations and the jets of active galactic nuclei are all confirmed sources of cosmic ray acceleration. For a review, see the recent \cite{Anchordoqui:2018qom}). 

The maximal cosmic ray energy that can be attained through these processes is not clear yet. However, given their observed gamma ray spectra it seems that, blazar jets \cite{1979ApJ...232...34B} produced by supermassive black holes (SMBHs) at the centre of galaxies  could be the most powerful accelerators in the Universe, with energies reaching far beyond the PeV scale  \cite{KRAWCZYNSKI2004367}. The discovery of (extragalactic) PeV neutrinos by the IceCube Neutrino Observatory \cite{Aartsen:2013jdh,Aartsen:2013bka} lends further support to the idea  that jets associated with SMBHs may be capable of accelerating particles above the PeV scale, although there is still debate on the primary composition of the of particles that are being accelerated (see e.g. \cite{2018arXiv181206025B} and references therein).  

Probing the particle content and the dynamics  of cosmic accelerators is not an easy task. Here, we argue that the detection of a gamma-ray circular polarisation signal from these sites could signal the presence of hadrons and hadronic collisions, independently of an observed neutrino signal.While leptonic processes such as synchrotron radiation also generate a circular polarisation depending on the orientation of the magnetic field, the polarisation analysed here is independent of the magnetic field. Our prediction is that hadronic collisions (e.g. proton-proton $pp$, proton-hadron and proton-photon $p\gamma$)  in cosmic accelerators or in the atmosphere (for a review see Ref.~\cite{Rieger:2013rwa}) produce photons with one dominant polarisation state and thus have the potential to create a net right-handed circular polarisation signal. Furthermore, this polarisation signal is quite frame and energy-independent for  proton-proton collisions but depends on the kinematics of the initial state for proton-photon collisions. Consequently, the polarisation also provide information on the dynamics of the distant accelerators.

This polarisation state is uncorrelated with the nature of the magnetic field in these acceleration sites (unlike synchrotron radiation), therefore the observation of a net circularly polarised $\gamma$-ray signal from a cosmic accelerator could reveal the intrinsic nature of its particle content. This conclusion holds whatever the astrophysical site under consideration as long as there is an excess of protons over antiprotons in the initial state. We also note that the interactions of protons in our galaxy and in the atmosphere will give a polarised photon background for new physics searches based on the polarisation of X-rays and gamma-rays~\cite{Boehm:2017nrl,Elagin:2017cgu}. Combined with newly-available neutrino data, the observation of a polarized signal would yield further information about the environment in which collisions are taking place, telling us for instance whether pion and muon decays are occurring in an optically thin or thick environment.

There are currently no planned experiments looking for circular polarisation of high energy gamma rays so far and the detection of such polarisation is experimentally challenging. However, we hope that the present results may further encourage development in this direction.

The paper is structured as follows: in Section~\ref{sec:analytic}, we study whether it is possible to generate a circular polarisation signal from the decay of Standard Model particles, using analytical arguments. In the same section, we also highlight the relationship between the neutrino flux  and the photons produced by these processes  \cite{PhysRevLett.66.2697}. As our analytical study shows that radiative decays could indeed generate circular polarisation signal,  we compute in Section~\ref{sec:prod} the fraction of circular polarisation expected from the decay of particles produced in proton-proton and photon-proton collisions for various toy configurations, including protons with TeV and PeV energies (and photons with TeV energies, e.g. from a prior CR collision) hitting a stationary target proton or a low energy photon, as expected when a relativistic jet interacts with the intergalactic medium. We also discuss head-on collisions of protons with centre of mass energies in excess of the Large Hadron Collider's, as could be expected inside astrophysical acceleration sites. The results for these various toy configurations give us some insights on which parameters are affecting the polarisation fraction. Therefore, they help us to understand both the more complicated and realistic case of proton spectrum and which processes can be distinguished thanks to circular polarisation. We discuss the results in Section~\ref{sec:discussion} and conclude in Section~\ref{sec:conclusion}.

\section{Photon and neutrino signals} 
\label{sec:analytic}

Circular polarisation arises from processes where the CP symmetry (the combination of the charge conjugation symmetry C and parity symmetry P) is violated, which generally occurs when the mechanism that produces the gamma-ray ray involves parity-sensitive couplings (i.e. a coupling proportional to the Dirac matrix $\gamma_5$) and a charge or particle-anti particle asymmetry. When these two conditions are met,  the photons are produced with one preferred polarisation and a net circular polarisation signal may  be observed from the site of production. We further note that the same electroweak processes which are the source of a net circular polarisation signal are also responsible for the generation of high energy neutrinos. Hence, there is a relationship between the fluxes of unpolarised photons and neutrinos which we will also study in this section.

\subsection{Radiative decay as a circular polarisation production mechanism} 

Hadronic collisions are known to produce hadrons, mesons, leptons and photons in profusion. The bulk of the photons thus generated originate from strong (non parity-violating) processes and therefore do not lead to the production of a circular polarisation signal. However a small fraction of the photons  produced in these collisions originate from weak, parity-violating processes such as the radiative decay of the neutron ($n$), charged pion ($\pi^\pm$), kaon ($K^\pm$) and muon ($\mu^\pm$)\footnote{We will neglect the contribution from the $\tau$ leptons as they are not abundantly produced in the processes considered here.}.  As proton-proton and proton-photon collisions are expected to produce an excess of $n$, $\pi^+$, $K^+$ and $\mu^+$ over $\bar{n}$, $\pi^-$, $K^-$ and $\mu^-$,  all the conditions are met for a circular polarisation signal to be generated\footnote{There are other P-violating weak sub-processes of the hadronic collision whose radiative corrections could lead to net photon polarisation, however there are subleading by many orders of magnitude. }.

Before investigating whether a charge asymmetry  can be generated in hadronic collisions, we start by verifying that such weak decays do generate photons with a preferred polarisation state. The main channel to produce photons from muon and meson decays is actually a radiative process\footnote{We will not discuss in this section the decays of the neutron and kaon but they follow the same pattern as described below.}.  Indeed, although the bulk ($> 99\%$) of the charged pions decay into $\pi^{\pm} \rightarrow \mu^{\pm} \, \nu_\mu$, a small fraction  (about 10$^{-4}$, though the exact value depends on the IR cutoff imposed on the gamma ray spectrum)  decays radiatively into $\pi^{\pm} \rightarrow \mu^{\pm} \, \nu_\mu \, \gamma$.  Similarly, the vast majority of muons  decay into $\mu^{\pm} \rightarrow e^{\pm} \, \nu_\mu \, \bar{\nu}_e$ but a small fraction ($\sim 10^{-3}$)  do decay radiatively into 
$\mu^{\pm} \rightarrow e^{\pm} \, \nu_\mu \, \bar{\nu}_e \, \gamma$.

The analytic expressions for these radiative decays were computed in \cite{Gabrielli:2005ek} and expressed for both $\mu^\pm$ and $\pi^\pm$  as the ratio of the differential rate to the first-order width\footnote{For the process $\pi^+ \rightarrow e^+ \nu \gamma$, $\Gamma_0$ would correspond to $\pi^+ \rightarrow e^+ \nu$, and not the full decay rate of the pion. } $\Gamma_0$. We recall these expressions in the next subsection.

\subsubsection{Pion radiative decay} 

The polarised pion decay $\pi^\pm \rightarrow \ell^\pm \nu_\mu \gamma$, with $\ell^\pm = \mu^\pm$ or $e^\pm$ takes the form \cite{Gabrielli:2005ek}:
\begin{equation}
\frac{1}{\Gamma_0}\frac{d\Gamma^{(\lell,\lga)}}{dxdy} = \frac{\alpha}{2\pi}\frac{1}{(1-r_\ell)^2}\rho^{(\lga,\lell)}(x,y)
\end{equation}
with
\begin{eqnarray}
\rho^{(\lga = \pm 1,\lell)}(x,y) &&= f_{IB}^{(\lga,\lell)}(x,y) \nonumber \\
&&+ \frac{\mpi^2}{f^2_\pi}\frac{(V\pm A)^2}{4 r_\ell}f_{SD}^{(\lga,\lell)}(x,y) \nonumber \\
&&+ \frac{\mpi}{f_\pi}\frac{(V\pm A)^2}{4 r_\mu}f_{INT}^{(\lga,\lell)}(x,y).
\label{eq:pidecay}
\end{eqnarray}
In these expressions $x = 2E_\gamma/\mpi$, $y = 2 E_e/\mpi$, $r_\ell \equiv m_\ell^2/\mpi^2$ and $\lga,\lell$ are the polarisation of the photon and lepton respectively  ($L = -1$ and $R = +1$) and $V$ ($A$) is the vector (axial) form factor of the pion, which we take from \cite{Patrignani:2016xqp}. In what follows, we will use $f_\pi = 0.131$ GeV. Three processes  contribute to this expression, namely 1) internal bremsstrahlung emission (IB) from either the initial or final state charged particle;  2) ``structure-dependent'' emission (SD) from the intermediate hadronic state; and 3) an interference term (INT) between the two former contributions. Their analytic forms are given in Appendix A of Ref.~\cite{Gabrielli:2005ek}.

 \subsubsection{Muon radiative decay} 

The radiative decay of the muon  $\mu^\pm \rightarrow e^\pm \nu_\mu \nu_e \gamma$ has the following form:
\begin{equation}
\frac{1}{\Gamma_0}\frac{d\Gamma^{(\lel,\lga)}}{dxdy} = \frac{\alpha}{24\pi} \frac{1}{A_ex}\left(G_0+\lga \bar G_0 + \lel(G_1 + \lga \bar G_1)\right)
\label{eq:mudecay}
\end{equation}
where $\lel$ and $\lga$ are respectively the electron and photon polarisations, $x = 2E_\gamma/\mmu$, $y = 2 E_e/\mmu$ and $A_e = \sqrt{y^2 - 4\me^2/\mmu^2}$. $G_{0,1}$ and $\bar G_{0,1}$ are polynomials in $x$ and $y$ which can be found in Appendix B of \cite{Gabrielli:2005ek}.

\subsection{Photon spectra \label{sec:polspectra}} 

The above expressions were given for pion and muon decaying at rest. However the mesons and leptons produced by hadronic collisions are not at rest, so photons must be successively boosted from the muon frame to the pion frame (if relevant), and subsequently to the ``lab'' frame of the observer. 

The spectrum from  the two-stage process $\pi \rightarrow \, \mu \, \nu$, $\mu \, \rightarrow \, e^- \, \nu_e \, \nu_{\mu} \, \gamma$ is given by 
\begin{equation}
\frac{d\phi_\gamma} {dE_\gamma} \sim \int d\Omega d E_{CM} \Lambda(E_{\gamma, \mathrm{lab}},E_{\gamma,CM})\frac{d\Gamma}{dE_{\gamma,CM}} \frac{d \phi_\pi}{dE_\pi}, 
\label{handwavy}
\end{equation}
where $\Lambda$ represents the boosts and rotations from the pion frame to the lab  frame, and ${d \phi_\pi}/{dE_\pi}$ is the pion spectral distribution in the lab frame. Eq. \eqref{handwavy} can equally be adapted to the muon decay spectrum, and sequential decays can be seen as nested integrals of the same form.

To explicitly evaluate the spectrum in the observer's frame, it is more practical to construct a simple Monte-Carlo simulation, which treats the differential decay rates as a probability distribution function (PDF). We first draw a photon with a random orientation, with an energy taken from the distribution given by \eqref{eq:pidecay} or \eqref{eq:mudecay}. This photon is then boosted to the lab frame and, by repeating the procedure for many photons, one eventually obtains the spectrum as seen by the observer. 

In the case of the two-step muon photons, the photon must first be boosted to the parent pion's frame, then boosted again to the lab frame, keeping track of the random direction of motion of the intermediate muon. We have verified that our results agree with the output of the effective models implemented in MadGraph5\_aMC@NLO. The advantage of this approach being faster sample generation and somewhat more transparent physics, thanks to the analytic form. We give the details of our prescription in Appendix \ref{app:MC}.

 Note that if we want the full spectrum down to $y= 2E_\gamma/m$, there is some subtlety: the cross section at leading order is IR-divergent. This divergence is canceled by the one-loop corrections to the polarised (non-bremsstrahlung) cross section. This is normally circumvented by including a cut in $E_{\gamma,min}$, corresponding to the minimum photon energy seen by the detector. We place this cutoff at $E_{min,\gamma} = 10$ MeV in the pion rest frame.
 \begin{figure}
\includegraphics[width=0.5\textwidth]{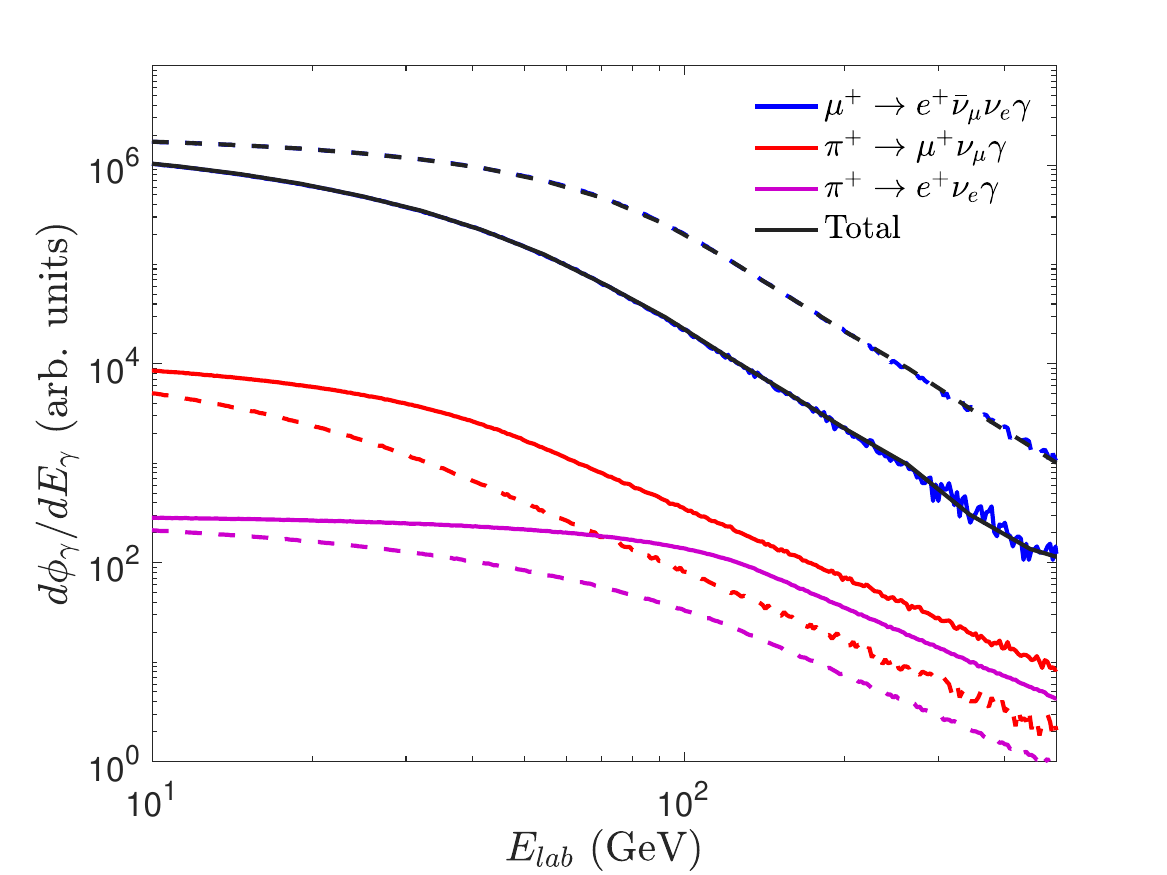}
\includegraphics[width=0.5\textwidth]{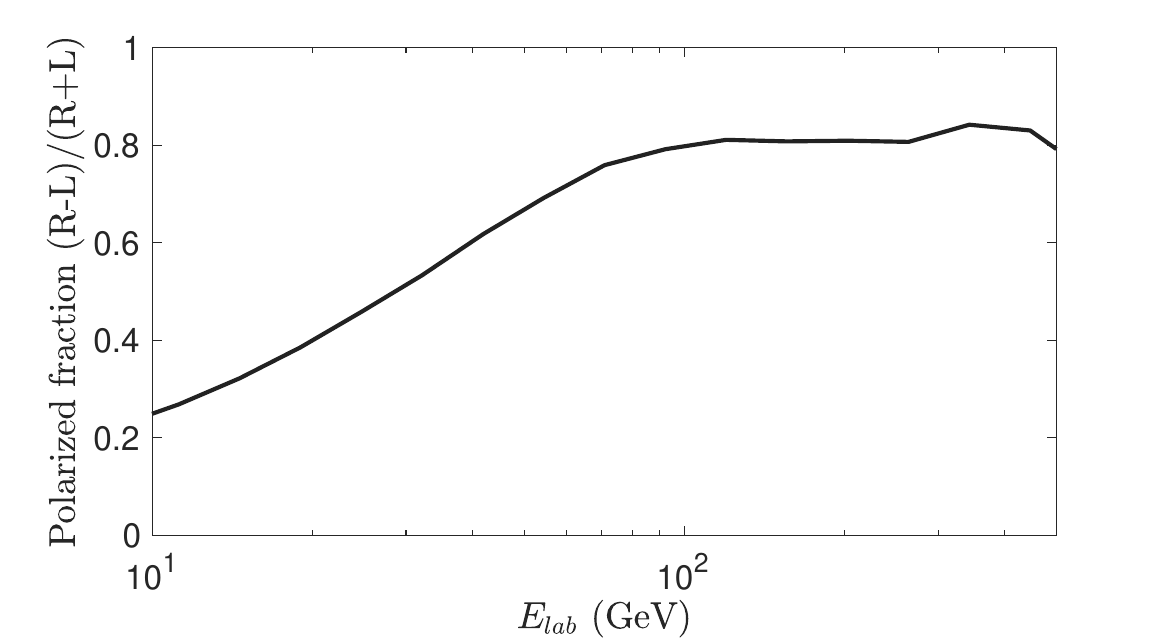}
\caption{polarised photons produced from the decay of a distribution of $\pi^+$, with a power law distribution in energies $d\phi_\pi/dE_\pi \propto E^{-\alpha}$, with $\alpha = 2.3$, and lower energy $E_{min} = 135$ GeV. Photon polarisations are represented by solid (left-handed) and dashed (right-handed) lines. Top: contributions to the total polarised flux; bottom: fractional contribution to the total flux of photons from charged pion decay from left ($P = -1$) and right-handed ($P = +1$) circularly polarised photons.}
\label{fig:specinlabframeg2}
\end{figure}

We present the flux of circularly polarised photons expected from pion and muon decays in Fig. \ref{fig:specinlabframeg2}, assuming a pion energy spectrum given by 
\begin{equation}
\frac{d\phi_\pi}{dE_\pi} \propto E^{-\alpha}.
\end{equation}

For concreteness we choose $\alpha = 2.3$, which coincides approximately with the expected pion spectrum from proton-proton collisions (see Sec. \ref{sec:prod}).

The  contribution to the radiative photon flux from the pion decay \eqref{eq:pidecay} into muons and electrons is shown in red and purple respectively. The radiated photon from muons is shown in blue. The total (black) overlaps with the blue line due to the overwhelmingly dominant contribution from muon decay. The dashed lines represent the right-handed (positive) polarised flux, while solid lines are left-handed (negative). 

The high-energy slope of the spectrum is determined by the spectral index of the pion flux, whereas the location of the break between the plateau and the slope is due to the minimum injected pion energy (135 GeV in this case). The bottom panel of Fig. \ref{fig:specinlabframeg2} shows the fractional polarisation of our signal. Above 50 GeV, about 80\% of the photons radiated from a positive pion have a right-handed polarisation.  Negative pion ($\pi^-$) decay would result in the exchange of L and R polarised photons. 

 Fig. \ref{fig:specinlabframeg2} shows that as long as there is some asymmetry between $\pi^+$ and $\pi^{-}$ production rates, there will be a net circular polarisation in CR acceleration sites.

 \subsection{High-energy neutrino signature}
\label{sec:HEnu}
In the previous subsection we argued that a net charge  asymmetry in the initial state together with parity violating interactions is at the origin of the final photon polarisation. Here we briefly turn our attention towards neutrinos. The relationship between neutrino flux and total gamma ray emission is well-known (see e.g. \cite{AlvarezMuniz:2002tn,Gaisser:2016uoy,Halzen2016}): gamma production in CR collisions comes dominantly comes from neutral pions, an the relationship between neutral and charged pion production guarantees such a correlation. Here, we show that this also leads to a relationship between the flux of neutrinos and the fluxes of \textit{polarised} photons.

The origin of this relationship stems in the fact that isospin is a good symmetry in a high energy $pp$ and $p\gamma$ interactions. That is, charged and neutral pions are expected to be produced equally frequently. Since neutral pions are the dominant source of unpolarised photons and charged pions are the dominant source of neutrinos, one expects the unpolarised photon and neutrino fluxes to be related, as discussed in . To demonstrate this explicitly we  consider the decay of a pion at rest into a final state that contains a particle $i$ and call $f_i(\Epsilon)$  the distribution function (which is related to the flux $d\phi_i/dE_i$) in terms of quantity $\Epsilon = E+p_z$ in the rest frame of the decaying pion. In the lab frame, boosted by $\gamma$, the energy of this particle is:
\begin{equation}
E_{\mathrm{lab}} = \gamma (E+ \beta p_z) \sim \gamma (E+p_z) = \gamma \Epsilon.
\end{equation}
The above expression assumes $\beta \sim 1$ because we are interested in highly boosted neutrinos that we can see in IceCube and other similar neutrino detectors.

Denoting $g_{\pi^0}(\gamma)$ the distribution function of the boosts of the neutral pions, the equation~\eqref{handwavy} can be simplified in terms of the boosts ($\gamma$) and $\Epsilon$ so that the distribution functions for the energies of particles $i$ in the lab frame (i.e. the differential flux (expected to be measured by the observer) reads:
\begin{align}
h_i(E_{\mathrm{lab}}) &= \int d\gamma \, d\Epsilon \, \delta( E_{\mathrm{lab}} - \gamma \Epsilon )  g_{\pi^0}(\gamma) \, f_i(\Epsilon).
\label{eq:hsubi}
\end{align}

The distribution functions $f_i(\Epsilon)$ can be evaluated analytically or using Monte Carlo methods as in the previous section. In the case $\pi^0 \to \gamma\gamma$, $f_
\gamma$ is particularly simple:
\begin{equation}
f_{\gamma}(\Epsilon) = \begin{cases} 
      \frac{2}{m_{\pi^0}} & 0 < \Epsilon < m_{\pi^0}, \\
	  0 & \mathrm{otherwise.}
   \end{cases}
\end{equation}
The normalisation of the above distribution is equal to 2, because there are two photons per neutral pion. As a result, the $h_\gamma$ from Eq:~\eqref{eq:hsubi} can be evaluated analytically using 
\begin{align}
h_{\gamma}(E_{\mathrm{lab}}) &= \frac{2}{m_{\pi^0}}\int_{E_{\mathrm{lab}}/m_{\pi^0}}^\infty \frac{d\gamma}{\gamma} \,  g_{\pi^0}(\gamma).
\end{align}
The fundamental theorem of calculus then guarantees that:
\begin{align}
g_{\pi^0}(\gamma) &=-\gamma \frac{m_{\pi^0}^2}{2} \left. \frac{dh_{\gamma}}{dE_{\mathrm{lab}}}\right|_{E_{\mathrm{lab}} = \gamma m_\pi}.
\label{eq:gpi}
\end{align}
The boost distribution $g_{\pi^0}(\gamma)$ for the neutral pions can then be derived using Eq.\ref{eq:gpi} from the measured photon spectrum ($h_{\gamma}$ assuming that all the photons come from the decay of neutral pions). Once $g_{\pi^0}(\gamma)$  is known, one can easily compute the expected neutrino spectra using Eq.~\eqref{eq:hsubi} and assuming $g_{\pi^+} = g_{\pi^-} = g_{\pi^0}$.

\begin{figure}
\includegraphics[width=0.45\textwidth]{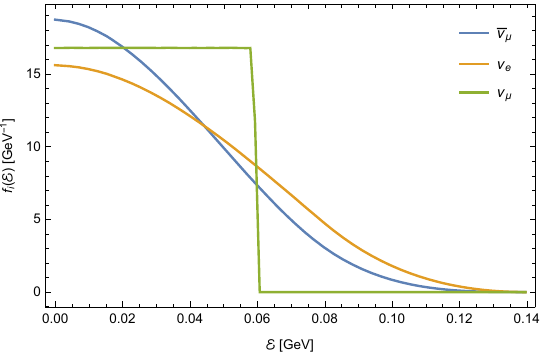}
\caption{Distributions $f_i(\Epsilon)$, where $\Epsilon = E+p_z$, for different decay products of $\pi^+$ in the decay channel $\pi^+ \to \mu^+ \nu_\mu \to e^+ \nu_
\mu \bar{\nu}_\mu \nu_e$. The distributions for decays of $\pi^-$ can be obtained by charge conjugation.}
\label{fig:gis}
\end{figure}

\begin{figure}
\includegraphics[width=0.45\textwidth]{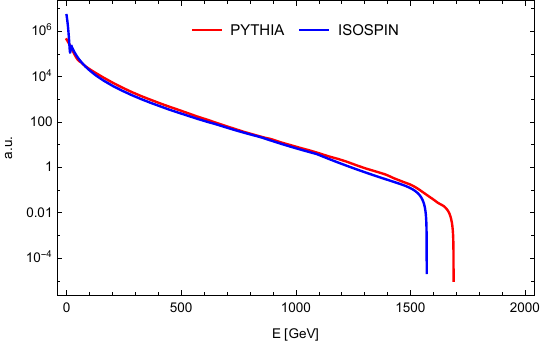}\\
\includegraphics[width=0.45\textwidth]{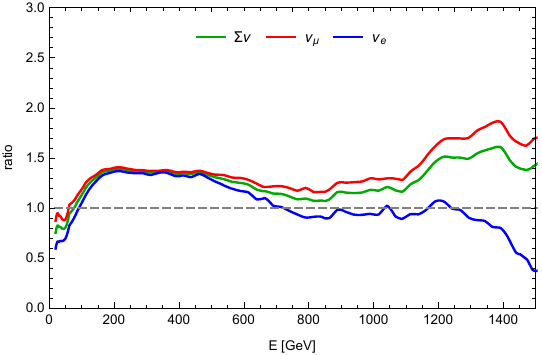}
\caption{The top figure shows the comparison between the neutrino spectrum computed by Pythia and the neutrino flux predicted from the photon flux using the method of section~\ref{sec:HEnu} from $pp$ collisions at 2 TeV. The bottom figure  shows the ratios for different neutrino species (where the sum over neutrino and anti-neutrino species is implied). Note that the isospin method is not reliable at high energy.}
\label{fig:nupred}
\end{figure}

The results are shown in Figure~\ref{fig:nupred} (top panel) where we display the ratio of the neutrino flux obtained directly from Pythia to the neutrino flux derived from the Pythia photon flux. 
As one can see the agreement is excellent and so we can be confident in our simulations\footnote{We have calculated the functions $f_{\nu_j}$ by numerically integrating over the full matrix element for the decay, and  show them in Figure~\ref{fig:gis}. }. We note that the photon flux extends to higher energies than the neutrino flux. This is expected since neutral pion decays into two photons, while the charged pion eventually produces four stable particles, reducing the  available energy per neutrino.  The kinematics associated with these two processes being different, the photon flux cannot be used to predict the neutrino flux up to the same energy; hence the mismatch between the cut-off locations in Figure~\ref{fig:nupred}.

\begin{figure}
\includegraphics[width=0.45\textwidth]{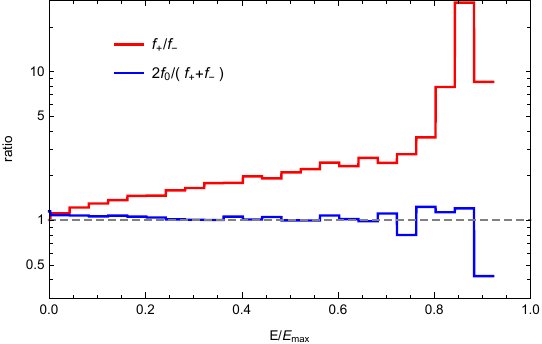}
\caption{Ratios of pion energy distribution functions  for 500 TeV protons impinging on protons at rest. Note that while the charge asymmetry grows steadily for larger energies, the neutral to charged pion flux stays much more stable.}
\label{fig:pionratios}
\end{figure}

The accuracy of the relationship $g_{\pi^+} + g_{\pi^-} \simeq 2g_{\pi^0}$ can be seen from the blue line in Figure~\ref{fig:pionratios}, which shows the ratios of pion distributions resulting from $pp$ collisions at 500 TeV. This ensures a relationship between the photon and neutrino fluxes. 

Conversely, the fact that $g_{\pi^+} \neq g_{\pi^-}$ (red line in Figure~\ref{fig:pionratios}) leads to our net polarisation signal as well as a neutrino-antineutrino asymmetry. This is detectable in principle, since the $\nu$-nucleon and $\bar \nu$-nucleon cross sections differ by a small amount. However, astrophysical and detector uncertainties make this extremely difficult.  The difference could be   noticeable in the case of electron neutrinos thanks to the Glashow resonance at $E_{\bar\nu_e} = 6.3$ PeV. 

\section{Net circular polarisation \label{sec:prod}}

Our analytical calculations have already indicated that the radiative decays of pions and muons should generate photons with a preferred polarisation state if there is a charge asymmetry in the initial state. We now show that both inclusive $pp$ and $p\gamma$ collisions can generate the required charge asymmetry (i.e. an excess of $\mu^+$ over $\mu^-$ etc) to eventually lead to a net circular polarisation signal in cosmic accelerators.

\subsection{Schematic procedure} 

We use the Pythia8 software \cite{Sjostrand:2006za,Sjostrand:2007gs,Sjostrand:2014zea} to simulate the nature and energy of the particles produced by the $pp$ and $p\gamma$  collisions.  The  photon spectra on which we base our conclusions are obtained in a 3-steps procedure summarised below. 

\begin{enumerate}
\item All the products of the collisions -- except for neutrons, charged pions, charged kaons and muons -- are allowed to decay. This is achieved by artificially setting the distance parameter in Pythia (which in the LHC context represents the distance between the ``detector'' and the collision) to $10^{18}$ mm\footnote{For consistency, we checked that the photons thus obtained were produced by parity conserving processes.} since the observation is expected to happen very far from the interaction point in astrophysics, contrary to colliders. This parameter should also be changed if interactions of CR in the atmosphere are considered instead.

\item The neutron, charged pion,  and muon decays are computed using the MadGraph5\_aMC@NLO \cite{Alwall:2011uj,Alwall:2014hca}.  
The pions and muons produced by the decay of the kaons are then added to their respective spectra. 

\item Finally we simulate the pion and muon decays using MadGraph5\_aMC@NLO to obtain the polarised photon spectra.
\end{enumerate}
 We do not keep the spin information from the first two steps. This could slightly affect our estimates of the polarisation fraction but this is not expected to have a major impact on the final results. In addition Pythia8 has not been calibrated  beyond a few TeVs yet (in the center-of-mass energy). Therefore, the normalisation of the hadronic spectra may not be exact at very high energy. In practice this corresponds to cosmic ray energies above a few tens of PeV. Finally, we mention that the center of mass energies in all the considered cases are well above the resonances of the $p-p$ and $p-\gamma$ cross-sections, as Pythia does not include such processes. These do not contribute to the $p-p$ collisions at these energies; for $p-\gamma$ collisions, we do note that resonances could increase the production of charged pions.

\subsection{Set-up}

We use the same parametrisation of Pythia8 as the ATLAS collaboration \cite{Aaboud:2016mmw} in their recent measurement of the inelastic $pp$ cross section. That is, we use the Monash \cite{Skands:2014pea} set of tuned parameters with the NNPDF 2.3 LO PDF and the pomeron flux model of Donnachie and Landshoff (DL) \cite{Donnachie:1984xq}. The parameters describing the pomoron Regge trajectory are set to $\alpha'=0.25$ and $\epsilon=0.1$. This configuration gives an inelastic cross-section of 78.4 mb at 13 TeV consistently with the ATLAS measurement. 

The output from Pythia (step 1) only contains stable particles (photons, electrons, protons, neutrinos) plus the neutrons, charged pions, kaons and muons  for which the decay has been switched off. The spectra for these four sets of particles (neutrons, charged pions, kaons and muons)  are displayed in Figures~\ref{fig:pp2T} and ~\ref{fig:pp2P} as solid lines,  for four scenarios: 1) a collision of a 2 TeV proton on a proton at rest, and 2) a collision of a 2 PeV proton on a proton at rest 3) a collision of two 6.5 TeV protons in the center of mass frame and 4) a collision of two 500 TeV protons in the centre of mass frame. The latter two scenarios are not expected to play a role in astrophysics. However, they serve to illustrate how frame and energy-independent our result is. 

The photons produced in the Pythia's output are unpolarised since they  come from the decay of mesons into two photons or the decay of excited QCD states into lower QCD states   without changing the flavour of the constituent quarks and neither processes can induce polarised photons (the initial state of the former is parity-even and the latter is dominated by the strong and electromagnetic interactions, which conserve parity). 

The pion decay is induced by the coupling of the $W$ boson to the leading term of the left-handed current of the chiral perturbation Lagrangian:
\begin{equation}
\frac{g_w f_\pi}{2\sqrt2} V_{ud}\partial_\mu \pi^- W^\mu
\label{effpion}
\end{equation}
where $g_w$ is the weak coupling constant, $f_\pi$ is the pion decay constant and $V_{ud}$ is the CKM matrix element. The neutron decay is induced by the following interaction:
\begin{equation}
g_n \bar{p}\gamma^\mu (1+r_{AV}\gamma_5) n W_\mu
\label{effneutron}
\end{equation}
where the overall coupling $g_n$ is fixed to give the measured total decay width of the neutron and the ratio of the axial and vector couplings is fixed at its PDG value  $r_{AV}=-1.2723$~\cite{Patrignani:2016xqp}.

The decay of the neutrons, charged pions, kaons and muons are computed in a second step using MadGraph5\_aMC@NLO and the low-energy models (including the effective interactions Eq.~\eqref{effpion} and Eq.~\eqref{effneutron}) implemented in FeynRules~\cite{Alloul:2013bka,Degrande:2011ua}. We start with the kaon decay as they produce more muons and pions through $K^+ \rightarrow \mu^+ \nu_\mu$, $K^+ \rightarrow \pi^0\pi^+$ and $K^+ \rightarrow \gamma \, \mu^+\nu_\mu$ (and charge-conjugate processes). We do not need to include the radiative corrections to $K$ decays into pions. These branching ratios are small and no net photon polarisation can be induced in these processes. The resulting charged pions and muons are added to the spectra obtained from Pythia and later decayed into: $\pi^+ \rightarrow \mu^+ \nu_\mu$, $\pi^+ \rightarrow \gamma \, \mu^+ \nu_\mu$, $\mu^+ \rightarrow e^+ \nu_\mu \bar \nu_e$ and $\mu^+ \rightarrow \gamma e^+ \nu_\mu \bar \nu_e$  respectively. The photons from the decay of the neutral pions are added to the unpolarised photon spectrum obtained in step 1. We have not included the neutrinos produced by the neutron decays ($n \rightarrow p e^- \nu$ and $n \rightarrow p e^- \nu \gamma$) to the initial neutrino spectrum produced by Pythia  as these  decays are very strongly phase-space suppressed.

Note that the convolution between the events from Pythia and the decay by Madgraph5\_aMC@NLO is not done event-by-event but bin-by-bin using the bin centre as the initial energy of every particle within each bin. We have checked however that a reduction of the binning by a factor two gives the same result within the numerical accuracy resulting from the limited number of events.

\subsection{Polarisation fraction from inclusive $pp$ collisions}

\begin{figure*}
\includegraphics[width=\textwidth]{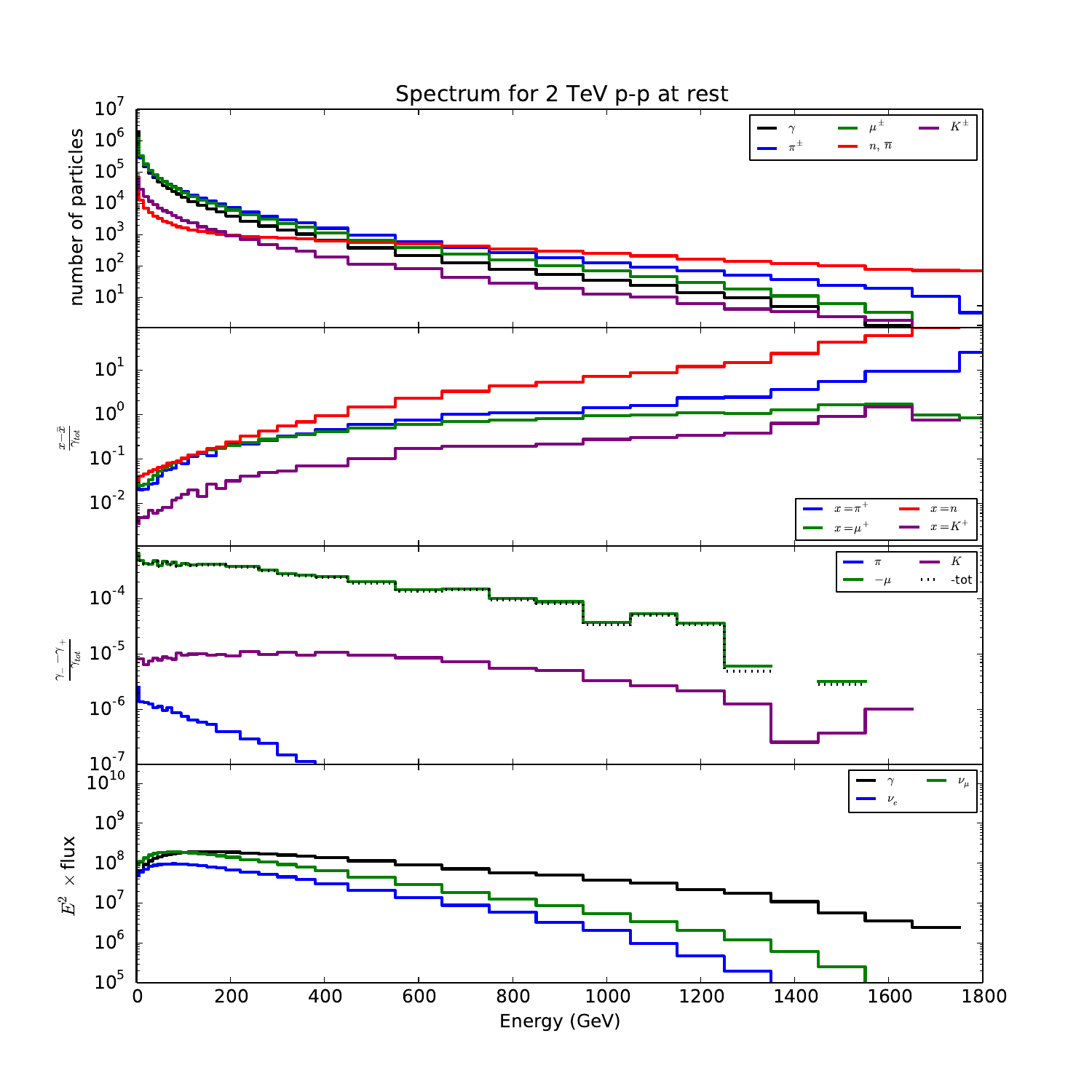}
\caption{Photon, charged pion, charged kaon, neutron, muon and neutrino spectra for 2 TeV protons hitting protons at rest, based on 200,000 events. The first panel show the total spectrum of the particles from Pythia  but also from the charged pions and kaons decays for the pions and muons and the photons of each polarisation from the charged pion, charged kaon, neutron and muon decays. The only cut on the photon is on its energy: $E_\gamma>0.01$GeV. The second panel displays the particle/anti-particle asymmetry relative to the photon spectrum. The polarisation fraction is shown on the third panel. In the muon case, the plotted quantity is $\gamma_+ - \gamma_-$ since $\mu^+$ produce dominantly right-handed photons. The neutrino spectrum is displayed in the last panel.}
\label{fig:pp2T}
\end{figure*}

\begin{figure*}
\includegraphics[width=\textwidth]{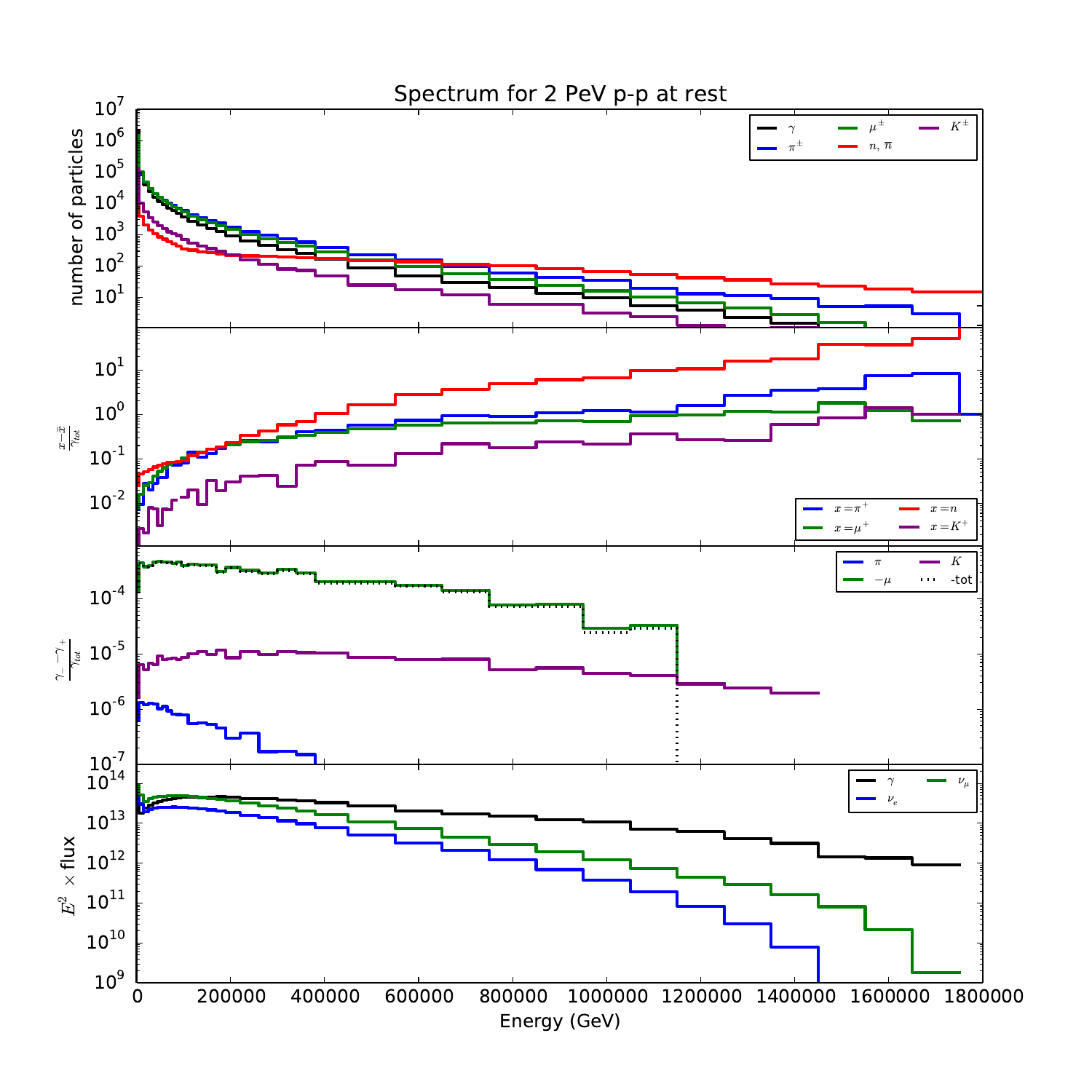}
\caption{Same as Fig.~\ref{fig:pp2T} for 2 PeV protons hitting protons at rest. We have not included the effects of propagation which lead to attenuation and appearance of an energy cut-off.}
\label{fig:pp2P}
\end{figure*}

\begin{figure*}
\includegraphics[width=\textwidth]{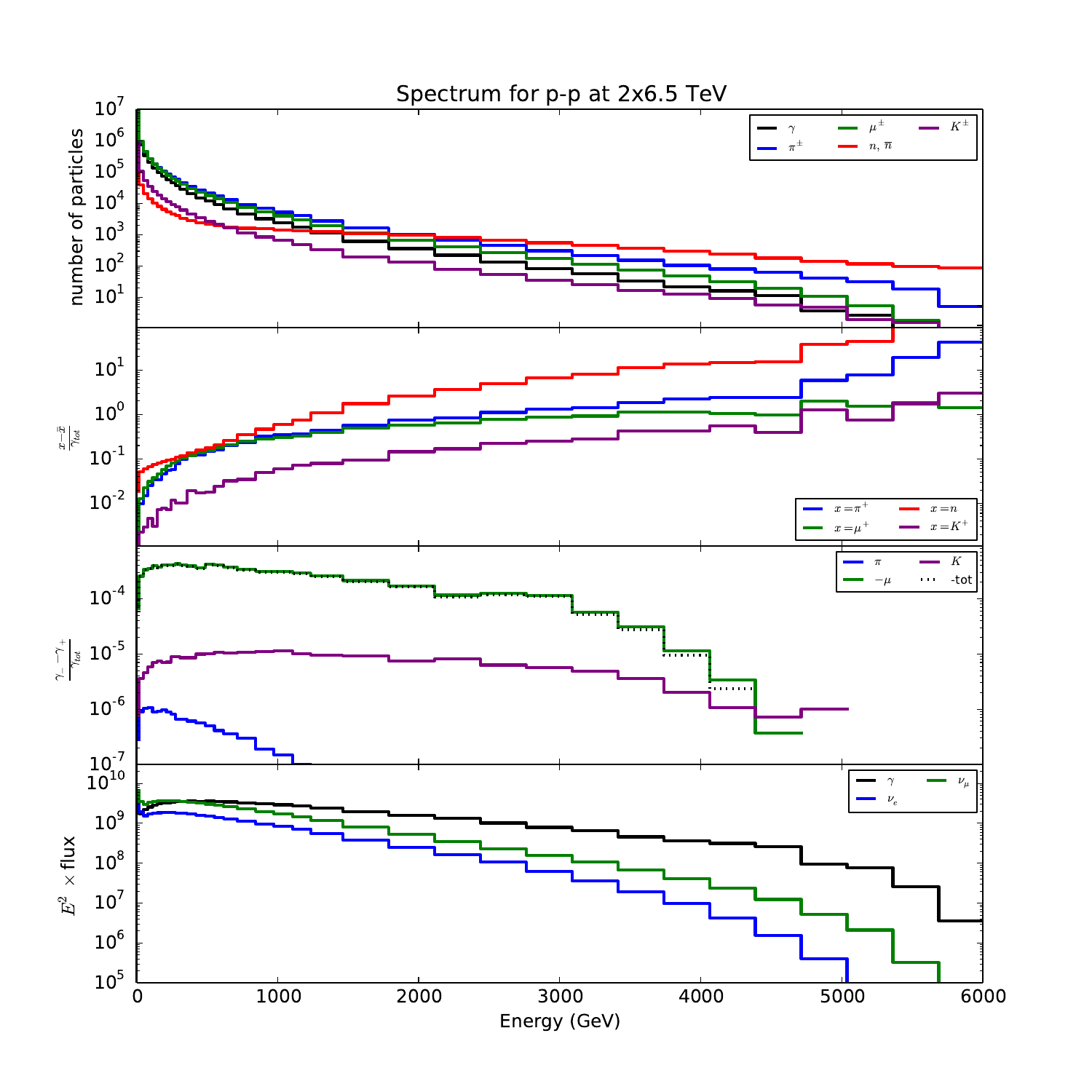}
\caption{Same as Fig.~\ref{fig:pp2T} for proton collision at 13TeV, inthe centre of mass (CM) frame. We have not included the effects of propagation which lead to attenuation and appearance of an energy cut-off.}
\label{fig:pp13TCM}
\end{figure*}

\begin{figure*}
\includegraphics[width=\textwidth]{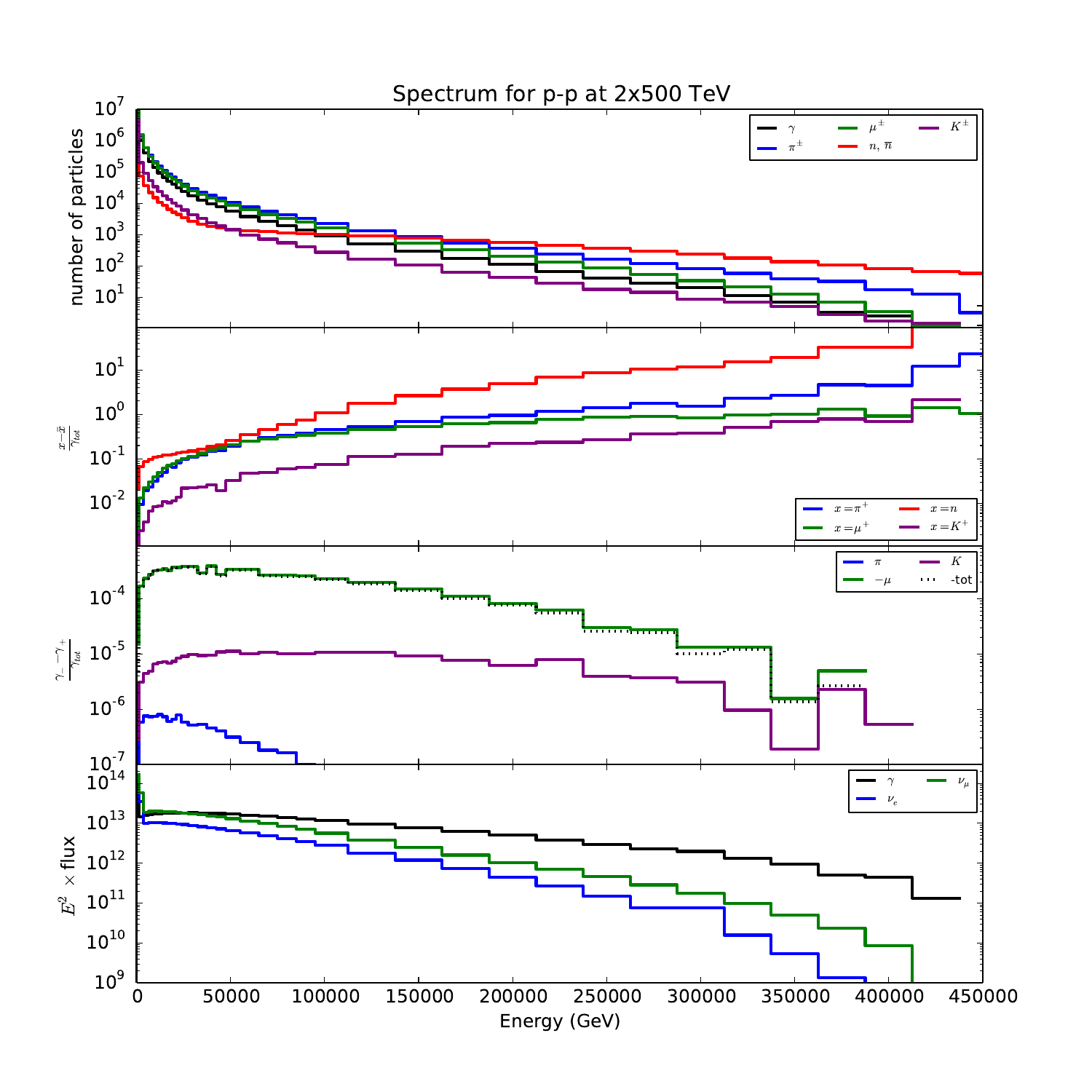}
\caption{Same as Fig.~\ref{fig:pp2T} for proton collision at 1PeV, in the centre of mass (CM) frame. We have not included the effects of propagation which lead to attenuation and appearance of an energy cut-off.}
\label{fig:pp1PCM}
\end{figure*}

The net circular polarisation signal that is generated from $pp$ collisions is displayed in the third panel of each figure~\ref{fig:pp2T},~\ref{fig:pp2P},~\ref{fig:pp13TCM} and \ref{fig:pp1PCM}. All four cases display a very similar behaviour, \textit{i.e.} a polarisation fraction on the order of $10^{-3}$, that decreases very slowly with energy. The nonzero polarisation is due to the combination of an asymmetry between particles and anti-particles (as seen in the second panel), and the degree to which parity is violated in the decay processes.

While neutrons produce the largest circular polarisation signal, all the photons that they produce have a very low energy due to the small phase space of the decay. They therefore end up in the lowest energy bin. This explains why the dominant contribution to the polarisation fraction is from the muon decays. 


\begin{figure*}
\includegraphics[width=\textwidth]{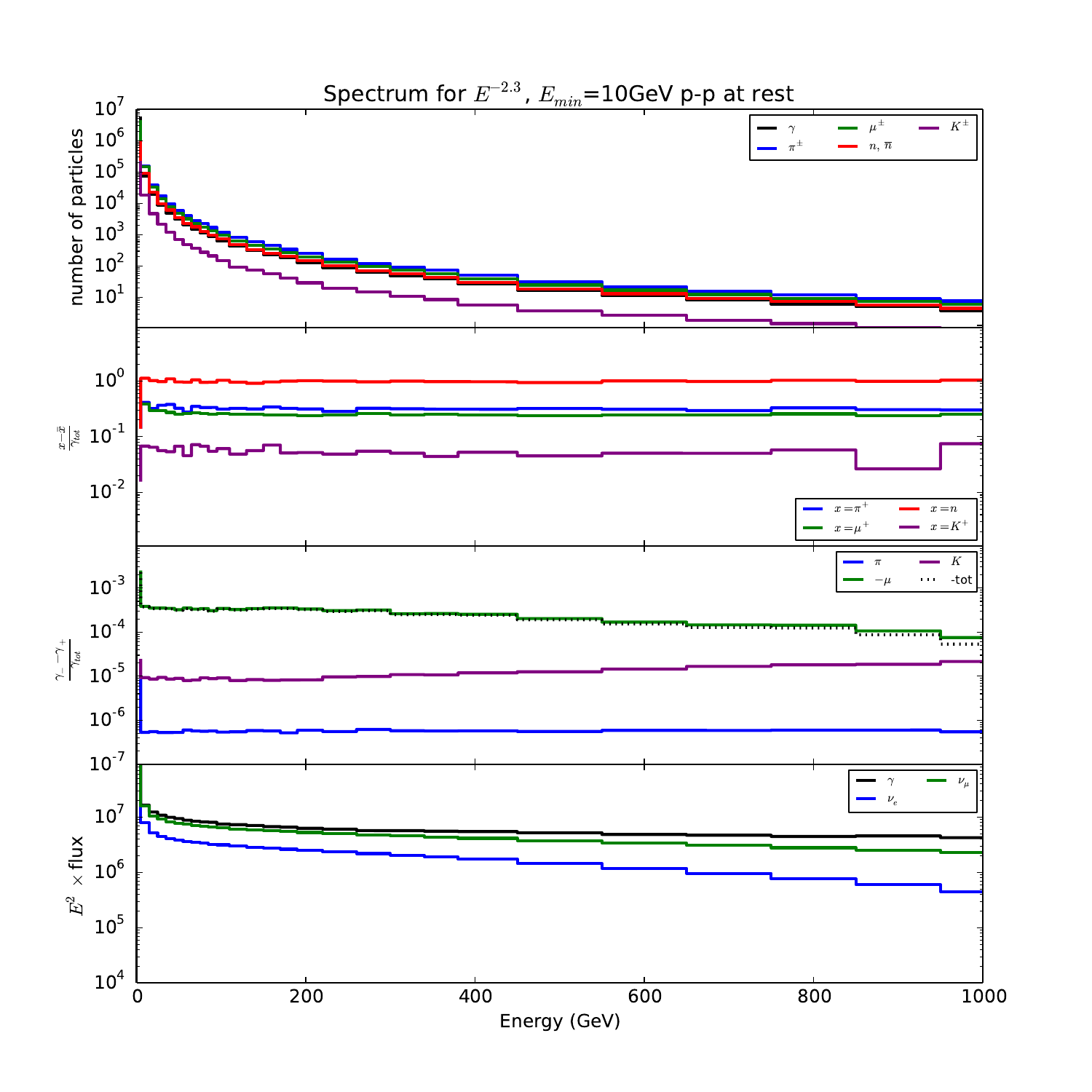}
\caption{Same as Fig.~\ref{fig:pp2T} but for protons with a power law spectrum $E^{-2.3}$ with a minimum energy of 10 GeV hitting protons at rest.}
\label{fig:ppdist}
\end{figure*}

Since the protons in a typical CR acceleration site are not monoenergetic but have a continuous energy distribution, we display  in Fig.~\ref{fig:ppdist} the result for collisions between protons at rest and a beam of  protons with a power law spectrum, \textit{i.e.} $\propto E^{-2.3}$ \cite{Castellina:2001ek}, with the minimal energy of the proton set at 10 GeV. The polarisation fraction is quite similar to the fixed energy case since it does not vary much with the initial energy of the protons or with the photon energies.  Given this similarity, we anticipate that changing the spectral index will have little effect on our conclusions.

Although the results of Pythia have been heavily compared to experimental data, the initial asymmetry in the pion and muon sector from $pp$ interactions has not  been experimentally confirmed yet. A few measurements have been made by CMS to compare the charged hadron and anti-hadron production~\cite{Sirunyan:2017zmn}. However the precision is insufficient  to confirm this effect. In addition, the CMS experiment focuses on low energy hadrons which contribute very little to the photon polarisation.  Therefore, our result should be seen as the best prediction that can be made so far.

\subsection{Polarisation fraction from inclusive  $p \gamma$}

\begin{figure*}
\includegraphics[width=\textwidth]{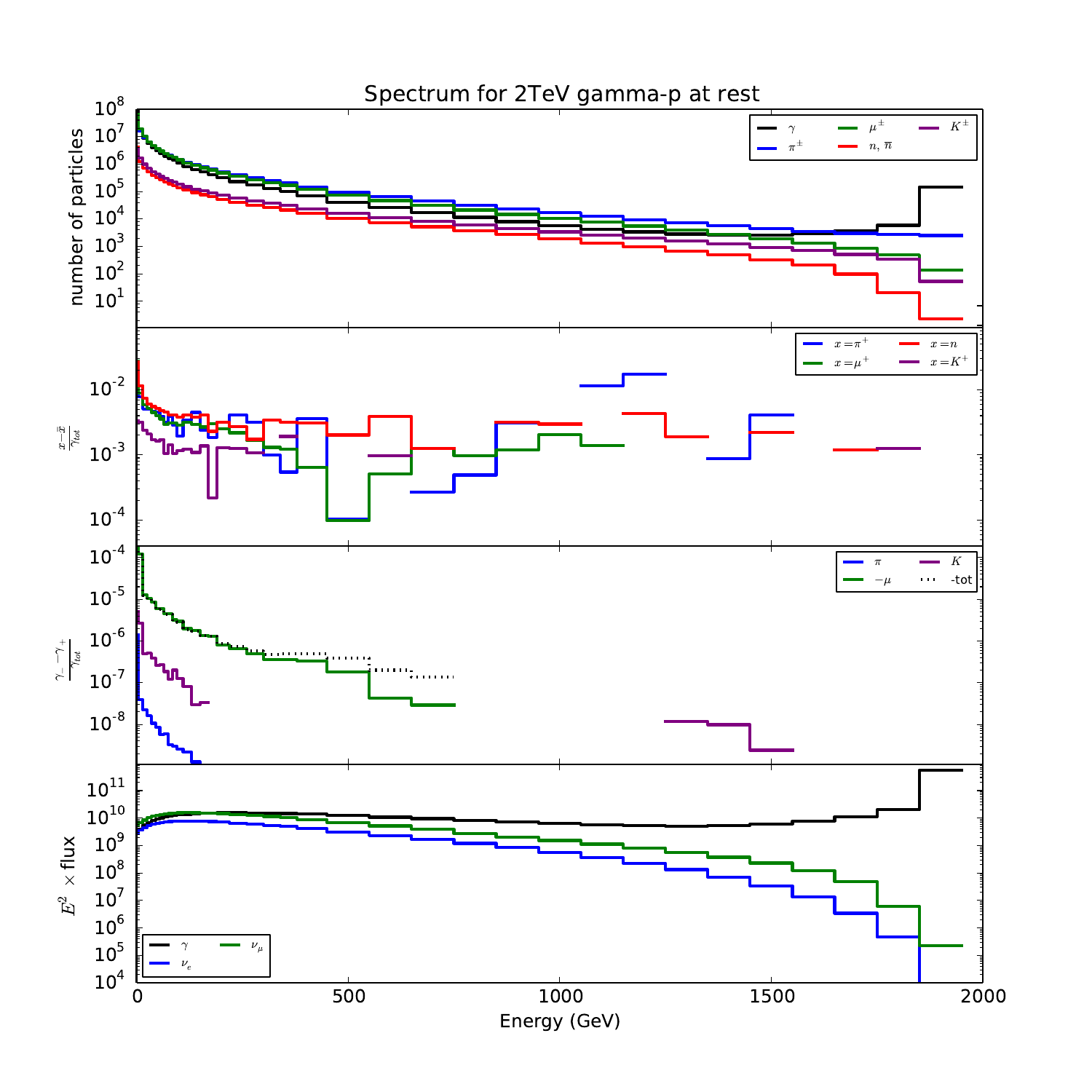}
\caption{Same as Fig.~\ref{fig:pp2T} for 2 TeV photons hitting protons at rest (Number of events is $10^7$). We have not included the effects of propagation which lead to attenuation and appearance of an energy cut-off.}
\label{fig:ap2T}
\end{figure*}

\begin{figure*}
\includegraphics[width=\textwidth]{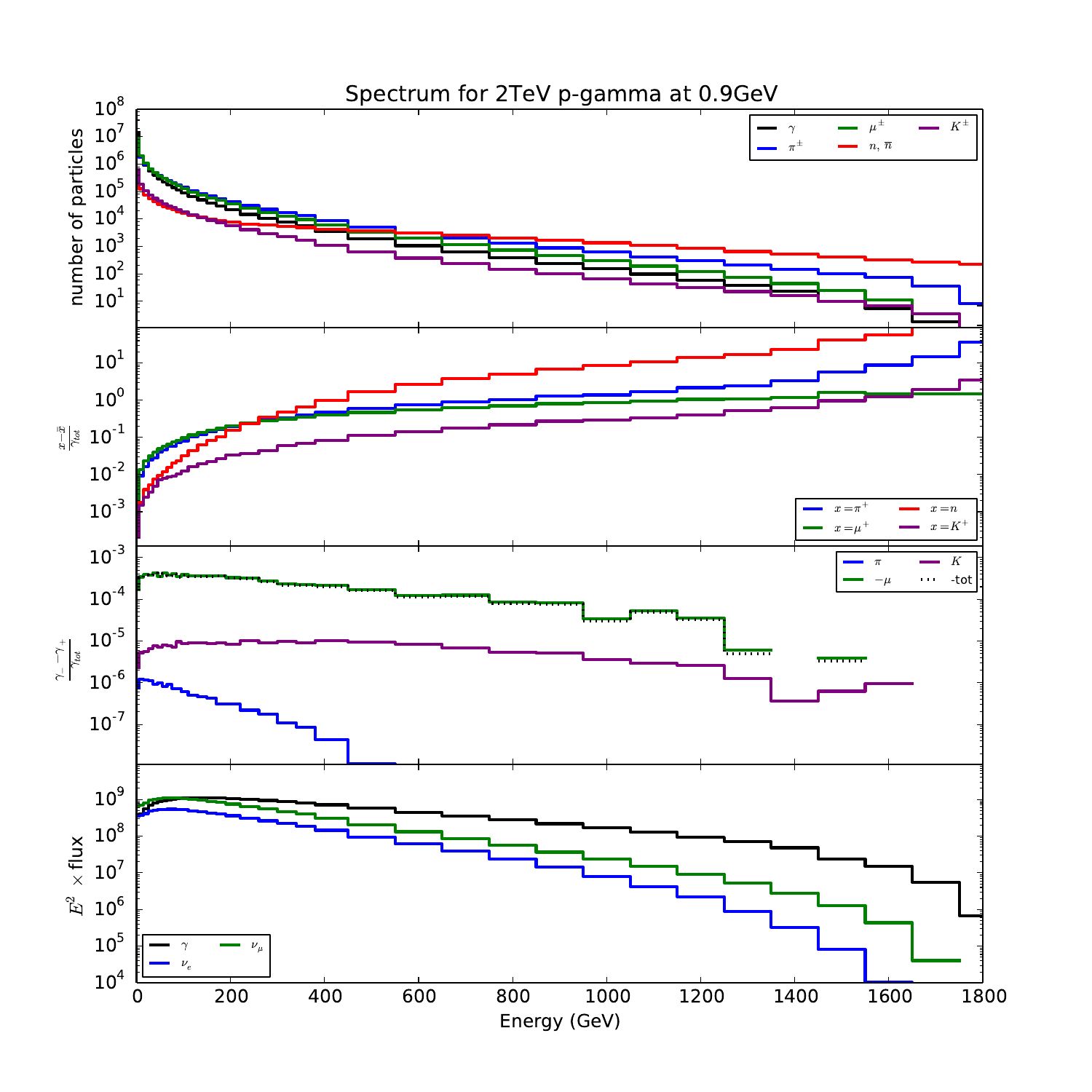}
\caption{Same as Fig.~\ref{fig:pp2T} for 2 TeV protons hitting low energy photons (Number of events is $10^6$). We have not included the effects of propagation which lead to attenuation and appearance of an energy cut-off.}
\label{fig:pa2T}
\end{figure*}

Proton-photon collisions are also simulated with Pythia~8, using the default setting which gives a total cross-section in agreement with H1 \cite{Aid:1995bz}. The procedure to obtain the photon polarisation is identical to the one used for proton-proton collisions. The results for collisions of 2 TeV photons on protons at rest are shown in Fig.~\ref{fig:ap2T}. Since SNRs such as the Crab are known to be high-energy gamma ray sources tat extend beyond 100 TeV \cite{Abeysekara:2019edl,Amenomori:2019rjd}, this could correspond to such a gamma hitting a proton from the interstellar medium. The asymmetry between particles and antiparticles, including the neutrinos, as well as the net photon circular polarisation, are a few orders of magnitude lower than for proton-proton collisions. 
This significant difference between $\gamma p$ and $pp$ collisions is due to the charge asymmetry of the proton remaining at rest and not being boosted to high energy like in the $pp$ collisions. Fig.~\ref{fig:pa2T} displays the opposite case: a high energy proton (with large charge momentum) hitting a low energy photon. In this case the polarisation fraction is similar to the $pp$ collisions. Although the total polarisation fraction is Lorentz invariant, the distributions and cuts are done in the observer frame and are at the origin of the difference between those two cases.   

\section{Discussion}
\label{sec:discussion}
In the above sections we have shown that hadronic processes in high-energy astrophysical environments can give rise to a non-negligible fraction of circularly polarised gamma rays. An observation of such a signal from a high-energy source would therefore help confirm the picture of a dominant hadronic component in AGN jets, and tell us whether neutrino production is occurring in an optically thin environment. Other scenarios can benefit from precision measurements of circular polarization: if the majority of gamma rays are produced by proton synchrotron, as in \cite{2019ApJ...876..109Z}, or if both hadrons and leptons contribute to the neutrino flux (e.g. \cite{Keivani_2018}), then a simple ratio of gamma ray-to-neutrino fluxes become insufficient. 

Depending on the measured value for the polarisation, we could furthermore discriminate between two scenarios based on the fact that  proton-proton collisions lead to a much stronger polarisation signal than collisions of high energy photons on stationary protons.

Furthermore, circular polarisation has recently been proposed as a signal of new physics \cite{Boehm:2017nrl,Elagin:2017cgu}. The polarisation produced by electroweak processes as described here is part of the background to such searches. Our result shows that this background is relatively small and almost independent of energy. Both are good news for new physics searches.

\subsection{Other sources of polarisation}
Some leptonic fraction in CR acceleration sites is inevitable:  synchrotron emission has been inferred from the linear polarisation of ultraviolet emission of at least one Blazar.  In an AGN jet, leptonic emission mechanisms at high energies fall into two flavours of inverse Compton scattering \cite{ghisellini1996origin}:
\begin{enumerate}
\item synchrotron self-Compton (SSC), wherein synchrotron emission is scattered to higher energies by a jet's relativistic electrons;
\item External Compton scattering (EC), either with the photons from the accretion disk, or with clouds external to the galaxy.
\end{enumerate}

Compton scattering leads to some suppression of the polarization fraction, but this amounts only to a few percent at the high energies $E_\gamma \gg m_e$ considered here \cite{Boehm:2019lvx}. 

Synchrotron radiation leads to both linear and circular polarisations depending on the relative direction between the magnetic field and the observer and is present also for lower energy photon. Therefore, circular polarisation from synchrotron radiation and hadronic processes can be distinguished according to their degree of correlation with the magnetic field, which can be inferred from the Faraday rotation of radio emission.

EC can yield GeV and higher photons; however, in the absence of a polarised electron flux, it would not yield a polarised gamma ray signal. If the jet magnetic field polarises the high-energy electrons, any asymmetry between $e^+$ and $e^-$ could plausibly lead to a polarised signal. Such an effect can be controlled by looking for a correlated neutrino signal, as we discussed in Section \ref{sec:HEnu}. 

Birefringence of the intergalactic medium induced by strong magnetic fields (the Cotton-Mouton effect) can act as a source of circular polarisation; however, this requires very strong magnetic fields ($\sim m_e^2/e$).

\subsection{Future prospects}
Measuring such a polarised gamma ray flux is obviously an observational challenge. Circular gamma ray polarisation has been measured in the laboratory \cite{Tashenov2011164,Tashenov2014,PhysRev.109.1015} so prospects for applying such techniques to astronomy are not unreasonable.  We leave an exact characterization of the flux normalization and background contributions to future work, as this requires detailed modeling of specific CR acceleration sites, and the inclusion of distance-dependent propagation effects.

There is an interesting opportunity to verify this work prior to observing individual sources of high energy photons of cosmic origin. The atmosphere is also a target for high energy protons and the $pp$ collisions produce the same degree of photon polarisation. It might be possible to detect this polarisation in future experiments and verify some of the theoretical models that we have used. Calculations would need to be modified to account for heavier cosmic ray isotopes, as well as the finite muon lifetime. The latter would reduce, and above certain energies, flip the relative polarisation of the signal.

\section{Summary}
\label{sec:conclusion}
Many processes at the origin of cosmic rays, including hadronic processes such as proton-proton or proton-photon collisions are expected to happen much more often that their CP-conjugate process due to the asymmetry between matter and antimatter in the observable Universe. The matter-antimatter asymmetry of the initial state results in an asymmetry between particle and anti-particle produced by the hadronic shower. This fact combined with the presence of Parity violation in the SM implies that they generate an asymmetry in the two circular polarisations of the resulting photons. Here, we have computed the circular polarisation fraction of the photons produced by the radiative decays following those two processes, and showed that they are about $5\times10^{-4}$ for both proton-proton collisions or collisions of energetic protons on a low energy photons. However, the polarisation fraction drops by more than one order of magnitude if the photon is carrying most of the energy of the collision and the proton is at rest. 
Here, we have considered several extreme scenarios/processes to illustrate how the polarisation fraction vary. While only some of them are expected to give the dominant contributions for many astrophysical sources, other processes such as high energy photon collisions with protons at rest would only give a small contribution. This small  contribution may increase if less extreme kinematics are chosen (lower energy photon on low energy proton, etc.) and their polarisation fraction will be closer but still different from the $p-p$ case. Eventually, precise polarisation measurements in the far future will pinpoint the various contributions.
Therefore, our results provide a test of the dynamics of distant objects as the primary acceleration mechanism has never been directly tested. While polarised light can also be induced by other processes such as synchrotron radiation or Compton scattering, the processes that we have focused on also lead to a correlated neutrino flux which we have explicitly derived and is uncorrelated with the magnetic field of the source.

\section*{Acknowledgements}
We thank Christine Rasmussen and Peter Skands for their help with PYTHIA. We also thank Frank Krauss for useful discussions. The work of C. D. was supported by the Fund for Scientific Research F.N.R.S.through the F.6001.19 convention. J. S. is grateful for support from COFUND. A.C.V. is supported by the Arthur B. McDonald Canadian Astroparticle Physics Research Institute, the Canadian Foundation for Innovation and the Ontario Ministry of Economics Development, Job Creation and Trade (MEDJCT). Research at Perimeter Institute was supported by the Government of Canada through the Department of Innovation, Science, and Economic Development, and by the Province of Ontario through MEDJCT.
 
 \appendix
\section{Monte Carlo boost to lab frame}
\label{app:MC}
\subsection{Pion three-body decay}
We start with the three-body decay $\pi^\pm \rightarrow \mu^\pm \nu_\mu \gamma$. We are only interested in the photon, so the muon energy $y$ can be integrated over. We first randomly select a photon energy in the pion's rest frame, with a probability $P(E_\gamma) \propto dN_\gamma/dE_\gamma =  d\Gamma_\pi/(\Gamma_\pi d\Eg)$. Randomly selecting a value of $\cos \theta$ in the uniform interval $[-1, 1]$, and $\phi \in [0,2 \pi]$ fully specifies the photon's four-momentum vector:
\begin{equation}
k^\alpha \equiv (\Eg, \Eg \cos \phi \sin \theta,\Eg \sin \phi \sin \theta,\Eg \cos \theta).
\label{eq:kdef}
\end{equation}
The isotropy of the photon emission allows us to choose a coordinate system such that the pion is traveling in the $x^1$ direction. Then the photon's four-momentum in the lab frame is:
\begin{equation}
k^\alpha_{lab} = \Lambda^\alpha_\beta k^\beta.
\end{equation}
If we are looking at a single pion energy $E_\pi = \gamma m_\pi$, then $\Lambda$ is trivial:
\begin{equation}
\Lambda =  \begin{pmatrix} 
 \gamma &-\beta\gamma& 0  &0 \\
  -\beta\gamma& \gamma & 0  &0 \\
  0 & 0 & 1 & 0 \\
  0 & 0 & 0 & 1 
 \end{pmatrix}
\end{equation}
However, if the pions are rather distributed with a spectrum of energies, we must select a boost from that distribution. In doing this, we also need to remember that the higher-energy pions will decay slower by a factor of $1/\gamma$, due to time dilation. This is completely equivalent to a suppression in the spectrum. Thus, we pick $\gamma$ from a distribution
\begin{equation}
P(\gamma) \propto \frac{1}{\gamma} \frac{dN_\pi}{d \gamma}.
\end{equation}
Repeating this process produces the lab-frame distribution of energies ($k^0_{lab}$) and momenta. Binning these energies is equivalent to integrating over an isotropic power law spectrum.

\subsection{Two-step muon decay}
The next scenario is the two-body decay $\pi^\pm \rightarrow \mu^\pm \nu_\mu$, followed by $\mu^\pm \rightarrow \gamma e^\pm \nu_e \nu_\mu$. Again, we proceed in ``reverse'' order, first boosting the final-state photon from the muon frame to the parent pion frame, then to the lab frame following some initial distribution of pions. As before, the initial photon distribution is isotropic in its parent frame (assuming we don't know the direction of the muon's spin), so Eq. \eqref{eq:kdef} can be used to generate the initial photon four-momentum, but this time using the muon decay spectrum.

The photon's four-vector in the pion frame is: $k'^\alpha = \Lambda^\alpha_\beta k^\beta$, though this time the boost is simple, since the two-body pion decay produces monoenergetic muons, with: $\gamma = (m_\pi/m_\mu + m_\mu/m_\pi)/2 = 1.0417$.

Now, we don't know which direction the muon was emitted in, with respect to the pion's direction of travel. This rotation must be done in the frame of the pion, on $k'^\alpha$. We can either randomly generate a rotation matrix, or simply replace the spatial components of $k'$ with a randomly-oriented vector with the same norm, using the prescription above for selecting random angles on a sphere. Schematically, we denote this new vector
\begin{equation}
k''^\alpha = R_\beta^\alpha \Lambda^\beta_\nu k'^\nu,
\end{equation}
remembering that $R_0^\alpha = R_\beta^0 = 0$. 

The final step is to boost back to the lab frame, given a pion energy distribution. The rotation above has allowed us to specify the pion direction of travel. The distribution is as above, except that we also need to account for the time-dilated muon decay rate. Fortunately, the muon's rest frame is very close to the pion's (since $\gamma_\mu = 1.04 \sim 1$), so this is dominated by the differences between the lab frame and the pion frame. This just means that there are now two time dilation factors: one slowing down pion decay, and one further slowing down muon decay:
\begin{equation}
P(\gamma) \propto \frac{1}{\gamma^2} \frac{dN_\pi}{d \gamma}.
\end{equation}

\bibliographystyle{JHEP_pat}
\bibliography{polbib}

\providecommand{\href}[2]{#2}\begingroup\raggedright\begin{thebibliography}{10}

\bibitem{Anchordoqui:2018qom}
L.~A. Anchordoqui, {\it {Ultra-High-Energy Cosmic Rays}},
  \href{http://arxiv.org/abs/1807.09645}{{\tt arXiv:1807.09645}}.

\bibitem{1979ApJ...232...34B}
R.~D. {Blandford} and A.~{K{\"o}nigl}, {\it {Relativistic jets as compact radio
  sources}},  {\em \apj} {\bf 232} (1979) 34--48.

\bibitem{KRAWCZYNSKI2004367}
H.~Krawczynski, {\it Tev blazars -- observations and models},  {\em New
  Astronomy Reviews} {\bf 48} (2004) 367 -- 373. 2nd VERITAS Symposium on the
  Astrophysics of Extragalactic Sources.

\bibitem{Aartsen:2013jdh}
IceCube: M.~G. Aartsen {\em et.~al.}, {\it {Evidence for High-Energy
  Extraterrestrial Neutrinos at the IceCube Detector}},  {\em Science} {\bf
  342} (2013) 1242856, [\href{http://arxiv.org/abs/1311.5238}{{\tt
  arXiv:1311.5238}}].

\bibitem{Aartsen:2013bka}
IceCube: M.~G. Aartsen {\em et.~al.}, {\it {First observation of PeV-energy
  neutrinos with IceCube}},  {\em Phys. Rev. Lett.} {\bf 111} (2013) 021103,
  [\href{http://arxiv.org/abs/1304.5356}{{\tt arXiv:1304.5356}}].

\bibitem{2018arXiv181206025B}
R.~{Blandford}, D.~{Meier}, and A.~{Readhead}, {\it {Relativistic Jets in
  Active Galactic Nuclei}},  {\em arXiv e-prints} (2018)
  [\href{http://arxiv.org/abs/1812.06025}{{\tt arXiv:1812.06025}}].

\bibitem{Rieger:2013rwa}
F.~M. Rieger, E.~de~Ona-Wilhelmi, and F.~A. Aharonian, {\it {TeV Astronomy}},
  \href{http://arxiv.org/abs/1302.5603}{{\tt arXiv:1302.5603}}.

\bibitem{Boehm:2017nrl}
C.~Boehm, C.~Degrande, O.~Mattelaer, and A.~C. Vincent, {\it {Circular
  polarisation: a new probe of dark matter and neutrinos in the sky}},  {\em
  JCAP} {\bf 1705} (2017) 043, [\href{http://arxiv.org/abs/1701.02754}{{\tt
  arXiv:1701.02754}}].

\bibitem{Elagin:2017cgu}
A.~Elagin, J.~Kumar, P.~Sandick, and F.~Teng, {\it {Prospects for detecting a
  net photon circular polarization produced by decaying dark matter}},  {\em
  Phys. Rev.} {\bf D96} (2017) 096008,
  [\href{http://arxiv.org/abs/1709.03058}{{\tt arXiv:1709.03058}}].

\bibitem{PhysRevLett.66.2697}
F.~W. Stecker, C.~Done, M.~H. Salamon, and P.~Sommers, {\it High-energy
  neutrinos from active galactic nuclei},  {\em Phys. Rev. Lett.} {\bf 66}
  (1991) 2697--2700.

\bibitem{Gabrielli:2005ek}
E.~Gabrielli and L.~Trentadue, {\it {Light mesons and muon radiative decays and
  photon polarization asymmetry}},  {\em Nucl. Phys.} {\bf B792} (2008) 48--88,
  [\href{http://arxiv.org/abs/hep-ph/0507191}{{\tt hep-ph/0507191}}].

\bibitem{Patrignani:2016xqp}
Particle Data Group: C.~Patrignani {\em et.~al.}, {\it {Review of Particle
  Physics}},  {\em Chin. Phys.} {\bf C40} (2016) 100001.

\bibitem{AlvarezMuniz:2002tn}
J.~Alvarez-Muniz and F.~Halzen, {\it {Possible high-energy neutrinos from the
  cosmic accelerator RX J1713.7-3946}},  {\em Astrophys. J.} {\bf 576} (2002)
  L33--L36, [\href{http://arxiv.org/abs/astro-ph/0205408}{{\tt
  astro-ph/0205408}}].

\bibitem{Gaisser:2016uoy}
T.~K. Gaisser, R.~Engel, and E.~Resconi, {\em {Cosmic Rays and Particle
  Physics}}.
\newblock Cambridge University Press, 2016.

\bibitem{Halzen2016}
F.~Halzen, {\it {High-energy neutrino astrophysics}},  {\em Nature Physics}
  {\bf 13} (2016) 232.

\bibitem{Sjostrand:2006za}
T.~Sjostrand, S.~Mrenna, and P.~Z. Skands, {\it {PYTHIA 6.4 Physics and
  Manual}},  {\em JHEP} {\bf 05} (2006) 026,
  [\href{http://arxiv.org/abs/hep-ph/0603175}{{\tt hep-ph/0603175}}].

\bibitem{Sjostrand:2007gs}
T.~Sjostrand, S.~Mrenna, and P.~Z. Skands, {\it {A Brief Introduction to PYTHIA
  8.1}},  {\em Comput. Phys. Commun.} {\bf 178} (2008) 852--867,
  [\href{http://arxiv.org/abs/0710.3820}{{\tt arXiv:0710.3820}}].

\bibitem{Sjostrand:2014zea}
T.~Sjostrand, S.~Ask, {\em et.~al.}, {\it {An Introduction to PYTHIA 8.2}},
  {\em Comput. Phys. Commun.} {\bf 191} (2015) 159--177,
  [\href{http://arxiv.org/abs/1410.3012}{{\tt arXiv:1410.3012}}].

\bibitem{Alwall:2011uj}
J.~Alwall, M.~Herquet, F.~Maltoni, O.~Mattelaer, and T.~Stelzer, {\it {MadGraph
  5 : Going Beyond}},  {\em JHEP} {\bf 06} (2011) 128,
  [\href{http://arxiv.org/abs/1106.0522}{{\tt arXiv:1106.0522}}].

\bibitem{Alwall:2014hca}
J.~Alwall, R.~Frederix, {\em et.~al.}, {\it {The automated computation of
  tree-level and next-to-leading order differential cross sections, and their
  matching to parton shower simulations}},  {\em JHEP} {\bf 07} (2014) 079,
  [\href{http://arxiv.org/abs/1405.0301}{{\tt arXiv:1405.0301}}].

\bibitem{Aaboud:2016mmw}
ATLAS: M.~Aaboud {\em et.~al.}, {\it {Measurement of the Inelastic
  Proton-Proton Cross Section at $\sqrt{s} = 13$  TeV with the ATLAS
  Detector at the LHC}},  {\em Phys. Rev. Lett.} {\bf 117} (2016) 182002,
  [\href{http://arxiv.org/abs/1606.02625}{{\tt arXiv:1606.02625}}].

\bibitem{Skands:2014pea}
P.~Skands, S.~Carrazza, and J.~Rojo, {\it {Tuning PYTHIA 8.1: the Monash 2013
  Tune}},  {\em Eur. Phys. J.} {\bf C74} (2014) 3024,
  [\href{http://arxiv.org/abs/1404.5630}{{\tt arXiv:1404.5630}}].

\bibitem{Donnachie:1984xq}
A.~Donnachie and P.~V. Landshoff, {\it {Elastic Scattering and Diffraction
  Dissociation}},  {\em Nucl. Phys.} {\bf B244} (1984) 322. [,813(1984)].

\bibitem{Alloul:2013bka}
A.~Alloul, N.~D. Christensen, C.~Degrande, C.~Duhr, and B.~Fuks, {\it
  {FeynRules 2.0 - A complete toolbox for tree-level phenomenology}},  {\em
  Comput. Phys. Commun.} {\bf 185} (2014) 2250--2300,
  [\href{http://arxiv.org/abs/1310.1921}{{\tt arXiv:1310.1921}}].

\bibitem{Degrande:2011ua}
C.~Degrande, C.~Duhr, {\em et.~al.}, {\it {UFO - The Universal FeynRules
  Output}},  {\em Comput. Phys. Commun.} {\bf 183} (2012) 1201--1214,
  [\href{http://arxiv.org/abs/1108.2040}{{\tt arXiv:1108.2040}}].

\bibitem{Castellina:2001ek}
A.~Castellina, {\it {Cosmic ray composition and energy spectrum above 1-TeV:
  Direct and EAS measurements}},  {\em Nucl. Phys. Proc. Suppl.} {\bf 97}
  (2001) 35--47, [\href{http://arxiv.org/abs/astro-ph/0011221}{{\tt
  astro-ph/0011221}}]. [,35(2001)].

\bibitem{Sirunyan:2017zmn}
CMS: A.~M. Sirunyan {\em et.~al.}, {\it {Measurement of charged pion, kaon, and
  proton production in proton-proton collisions at $\sqrt{s}=13$ TeV}},  {\em
  Phys. Rev.} {\bf D96} (2017) 112003,
  [\href{http://arxiv.org/abs/1706.10194}{{\tt arXiv:1706.10194}}].

\bibitem{Aid:1995bz}
H1: S.~Aid {\em et.~al.}, {\it {Measurement of the total photon-proton
  cross-section and its decomposition at 200-GeV center-of-mass energy}},  {\em
  Z. Phys.} {\bf C69} (1995) 27--38,
  [\href{http://arxiv.org/abs/hep-ex/9509001}{{\tt hep-ex/9509001}}].

\bibitem{Abeysekara:2019edl}
HAWC: A.~U. Abeysekara {\em et.~al.}, {\it {Measurement of the Crab Nebula at
  the Highest Energies with HAWC}},
  \href{http://arxiv.org/abs/1905.12518}{{\tt arXiv:1905.12518}}.

\bibitem{Amenomori:2019rjd}
M.~Amenomori {\em et.~al.}, {\it {First Detection of Photons with Energy Beyond
  100 TeV from an Astrophysical Source}},
  \href{http://arxiv.org/abs/1906.05521}{{\tt arXiv:1906.05521}}.

\bibitem{2019ApJ...876..109Z}
H.~{Zhang}, K.~{Fang}, {\em et.~al.}, {\it {Probing the Emission Mechanism and
  Magnetic Field of Neutrino Blazars with Multiwavelength Polarization
  Signatures}},  {\em \apj} {\bf 876} (2019) 109,
  [\href{http://arxiv.org/abs/1903.01956}{{\tt arXiv:1903.01956}}].

\bibitem{Keivani_2018}
A.~Keivani, K.~Murase, {\em et.~al.}, {\it A multimessenger picture of the
  flaring blazar {TXS} 0506$+$056: Implications for high-energy neutrino
  emission and cosmic-ray acceleration},  {\em The Astrophysical Journal} {\bf
  864} (2018) 84.

\bibitem{ghisellini1996origin}
G.~Ghisellini and P.~Madau, {\it On the origin of the $\gamma$-ray emission in
  blazars},  {\em Monthly Notices of the Royal Astronomical Society} {\bf 280}
  (1996) 67--76.

\bibitem{Boehm:2019lvx}
C.~Boehm, A.~Olivares-Del~Campo, M.~Ramirez-Quezada, and Y.-L. Zhou, {\it
  {Polarisation of high energy gamma-rays after scattering}},
  \href{http://arxiv.org/abs/1903.11074}{{\tt arXiv:1903.11074}}.

\bibitem{Tashenov2011164}
S.~Tashenov, {\it Circular polarimetry with gamma-ray tracking detectors},
  {\em Nuclear Instruments and Methods in Physics Research Section A:
  Accelerators, Spectrometers, Detectors and Associated Equipment} {\bf 640}
  (2011) 164 -- 169.

\bibitem{Tashenov2014}
S.~Tashenov, T.~Back, {\em et.~al.}, {\it Electron polarimetry with
  bremsstrahlung},  {\em Journal of Physics: Conference Series} {\bf 488}
  012057.

\bibitem{PhysRev.109.1015}
M.~Goldhaber, L.~Grodzins, and A.~W. Sunyar, {\it Helicity of neutrinos},  {\em
  Phys. Rev.} {\bf 109} (1958) 1015--1017.

\end{thebibliography}\endgroup

\end{document}